\documentclass{kluwer}    % Specifies the document style.
\usepackage[dvips]{graphicx}

\newdisplay{guess}{Conjecture}

\begin{document}                                  
                                                 
\begin{article}
\begin{opening}         
\title{A fundamental plane of black hole activity: pushing forward the
  unification scheme} 
\author{Andrea \surname{Merloni}\email{am@mpa-garching.mpg.de}}
\institute{Max-Planck-Institut f\"ur Astrophysik,
 Garching, Germany}
\author{Sebastian \surname{Heinz}}
\institute{Center for Space Research, MIT; Chandra Fellow}
\author{Tiziana \surname{Di Matteo}}
\institute{Max-Planck-Institut f\"ur Astrophysik,
 Garching, Germany}
\runningauthor{A. Merloni, S. Heinz \& T. Di Matteo}
\runningtitle{A fundamental plane of black hole activity}

\date{October 20, 2004}

\begin{abstract}
We examine the disc--jet connection in stellar mass and supermassive
  black holes by investigating the properties of their compact emission in
  the hard X-ray and radio bands.  We compile a sample of $\sim$100 active
  galactic nuclei with measured mass, 5 GHz core emission, and 2-10 keV
  luminosity, together with 8 galactic black holes with a total of $\sim$
  50 simultaneous observations in the radio and X-ray bands. Using this
  sample, we study the correlations between the radio ($L_{\rm R}$) and
  the X-ray ($L_{\rm X}$) luminosity and the black hole mass ($M$).  We
  find that the radio luminosity is correlated with {\em both} $M$ and
  $L_{\rm X}$, at a highly significant level. We show how this result
  can be used to extend the standard unification by orientation scheme
  to encompass unification by mass and accretion rate.
\end{abstract}

\end{opening}           

\section{Introduction: unified pictures}  

Some galaxies are known to emit radiation with extremely high
luminosities in the $\gamma$-ray, X-ray, UV and radio continuum from a
very concentrated volume in the nuclear region. Such active cores are
the so-called Active Galactic Nuclei (AGN) and their radiation is
believed to be produced by accretion onto a supermassive black hole.

The intrinsically complex nature of such systems and the differences
in the terminology among different scientific communities (radio,
optical, X-ray astronomers) has led to an extremely complicated
nomenclature for the AGN zoo.  As the wealth of observations piled up,
and with them the number of different AGN types, the opposite
enterprise of finding unification schemes has progressively gained
support. The basic idea behind the standard unification scheme is that
AGN are asymmetric and anisotropic systems. This is natural, as all
rotating systems necessarily single out a preferential axis in space
and break the full spherical symmetry of non-rotating
bodies. Therefore, the orientation of the AGN rotation axis with
respect to our line of sight becomes another important parameter that
can cause apparent observational differences in two sources that are
intrinsically identical ({\it unification by orientation}).

According to the current widely accepted paradigm (Urry and Padovani
1995), there are two principal causes of anisotropic radiation:
obscuration and relativistic beaming.  The former is usually
associated with a torus of gas and dust obscuring the optical, UV and
(sometimes) soft X-ray radiation along some line of sights, the latter
with outflows of energetic particles (jets) along the symmetry axis:
the high velocity plasma in the jet beams radiation relativistically
in the forward direction.  The powerful source of radiation and the
launching site of the collimated jet lies in the center of the system.

Observationally, jet morphologies and spectral properties of both
radio and X-ray cores are remarkably similar in the case of black
holes of stellar mass (Galactic black holes, hereafter GBH) and of
their supermassive counterparts in the nuclei of galaxies (hereafter
SMBH).  If jets are launched in the innermost parts of the accretion
flows, as commonly assumed, then these similarities suggest that it
should be possible to extend the unification scheme further and to
understand the physics of both black hole accretion and jet production
by studying all those systems as a single class ({\it unification by
mass}).  Furthermore, the recent discovery that SMBH lie at the center
of the majority (possibly all) galactic nuclei, even in apparently
inactive ones (Kormendy and Richstone 1995; Magorrian et al. 1998),
naturally leads to the idea that a grand unification is possible by
taking into account the differences in fueling rates among different
objects ({\it unification by accretion rate}).

In the following, we will show how to proceed quantitatively towards a
grand unification of active black holes by studying the multivariate
correlation among masses, radio luminosities and hard X-ray
luminosities of objects traditionally classified in the more diverse
ways. The underlying theoretical assumption is that both accretion and
jet production are fundamental manifestation of black hole activity,
and are somehow physically connected (Begelman et al. 1984; Rawlings
and Saunders 1991; Falcke and Biermann 1995; Heinz and Sunyaev 2003).

\section{Unification by mass and accretion rate}

\begin{figure}
\includegraphics[angle=270,width=0.9\textwidth]{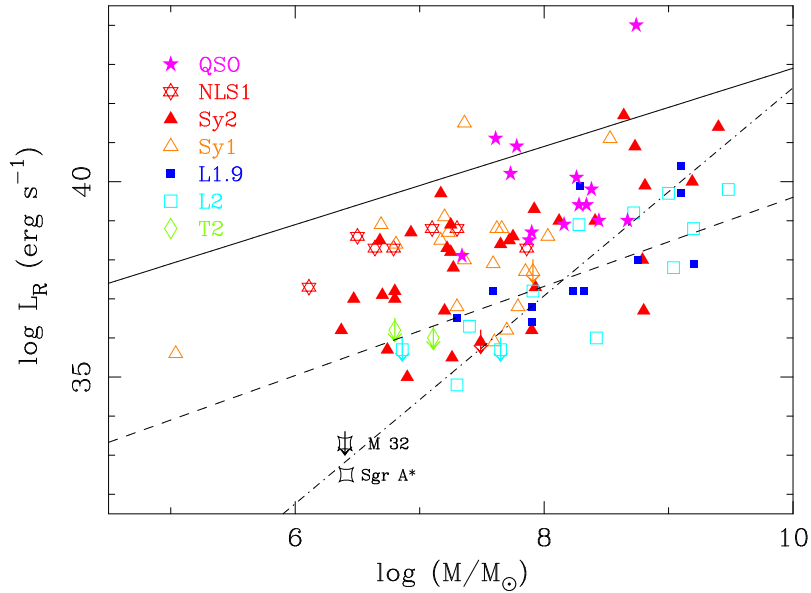}
\caption[]{Radio core luminosity at 5 GHz versus black hole mass of
  SMBH only. Upper
  limits are marked with arrows, different symbols indicate objects
  belonging to different spectral classes. 
  The dot-dashed line gives the regression
  fit proposed by Franceschini et al. (1998), the  dashed line
  that proposed by Nagar et al. (2002), both obtained using different
  samples of SMBH only. The thick solid upper line gives the maximum
  core radio power as calculated by Ho (2002) for sources accreting at
  the Eddington rate.}
\label{fig:lr_m2}
\end{figure}

Here we briefly describe the main results of the
correlation analysis carried on in
Merloni, Heinz and Di Matteo (2003, hereafter MHD03).
A more detailed description of the sample used, of the
selection effects and of the statistical analysis can be found there, together with a
comprehensive list of references to the observational data.
 For future reference, we define the dimensionless
black hole mass $M=M_{\rm BH}/M_{\odot}$ and accretion rate $\dot m \equiv
(L_{\rm bol}/\epsilon)/L_{\rm Edd} = \dot M c^2/L_{\rm Edd} \propto \dot M/M$,
where $\epsilon$ is the accretion efficiency.

We have selected from the existing literature a sample of black
hole-powered systems with measured masses, the nuclei of which have been
observed both at 5 GHz (mostly with arcsecond resolution with the VLA) and
in the 2-10 keV band. The main obvious advantage of this choice
lies in the fact that obscuration is unimportant, or
easily accounted for, in these spectral bands.  

We first considered the full sample of $\sim 40$ nearby inactive, or weakly
active galaxies with existing nuclear black hole mass measurements from
observations of spatially resolved kinematics. To these we have added a
comparable number of bright AGNs (and QSOs) with nuclear black hole mass
measured from reverberation mapping of their broad line region. 
From this sample we selected all
objects which have been observed in both the radio and X-ray bands.
In order to obtain a more statistically representative sample, we also
searched the existing literature for both nearby low-luminosity galactic
nuclei and for relatively bright Seyfert nuclei (either
type 1, type 2 or Narrow Line Seyfert 1) and radio galaxies with available
radio and X-ray flux measurements.  We assign black hole masses to these systems
using the observed correlation between black hole masses and stellar
velocity dispersion (Gebhardt et al. 2000; Ferrarese and Merritt 2000). 

Relativistically beamed sources (i.e. those whose jet axis points
towards our line of sight) are dominated by the boosted jet emission,
and cannot be used to test the disc-jet coupling. We therefore
excluded from our sample BL Lac objects. Among the Quasars in our
sample, only 3C 273, which has an extremely high radio loudness and a
blazar-like spectrum, is likely to suffer from strong Doppler boosting
of the radio jet. On the other hand, according to the unification scheme, 
Seyfert 2 nuclei should not be preferentially viewed pole on, while
for all the other sources (mainly low-luminosity AGN and Seyfert 1),
for which the nature of the (relatively faint) radio emission is not
well established, we have assumed that the
orientation of their jets with respect to line of sight is randomly
distributed.

\begin{figure}
\includegraphics[angle=270,width=0.9\textwidth]{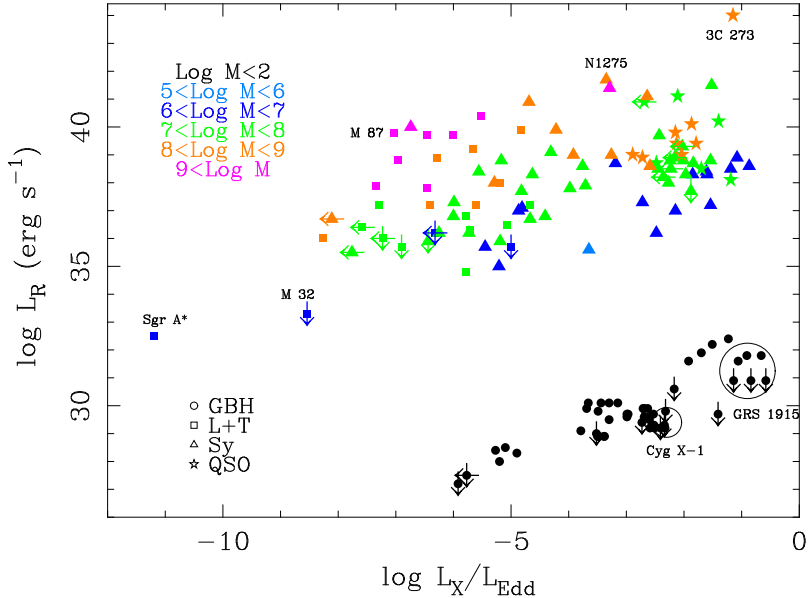}
\caption[]{Radio
  core luminosity at 5GHz vs. the ratio $L_{\rm X}/L_{\rm Edd}$ of
  X-ray to Eddington luminosity. Upper
  limits are marked with arrows, different symbols indicate objects
  belonging to different spectral classes and different colors objects
  in different mass bins. The color-coding of the different mass
  bins makes the mass segregation more evident.}
\label{fig:lr_mdot2}
\end{figure}

The Galactic X-ray binaries included in our sample have been selected
to have (a) simultaneous X-ray and radio observations, or {\em RXTE}
All-Sky-Monitor (ASM) X-ray data in conjunction with radio fluxes
available from the literature, and (b) publicly available {\em
RXTE}-ASM X-ray and Green-Bank Interferometer (GBI) radio lightcurves
(from which we estimated the 5 GHz fluxes by interpolating between the
2.25 GHz and the 8.3 GHz channels).  

The final sample consists of $\sim$100 active
  galactic nuclei with measured mass, 5 GHz core emission, and 2-10 keV
  luminosity, together with 8 galactic black holes with a total of $\sim$
  50 simultaneous observations in the radio and X-ray bands (see MHD03
  for a full list of the sample sources).

In  Figure~\ref{fig:lr_m2} we show the radio core luminosity versus the
black hole mass for objects of different spectral classes. We concentrate on the SMBH
only in order to show how, at a fixed black hole mass, sources which
are classically regarded as powerful accretors (QSOs, Narrow Line
Seyfert 1, Seyfert 1) tend to lie above the so-called Low-Luminosity AGN (here
represented by LINERs), with Seyfert 2 galaxies spanning a large area
in the vertical direction.
Figure~\ref{fig:lr_mdot2} shows instead the core radio luminosity versus
the ratio of the X-ray nuclear luminosity to the Eddington luminosity
(probably a good estimator of $\dot m$, see below).  
We represent objects in different mass bins with different
colors. It is clear that, when the
data points are grouped into mass bins, objects in different bins tend
to lie on parallel tracks. The presence of a mass
segregation suggests that the radio luminosity of an object likely
depends both on its accretion rate and on its mass.

\begin{center}
\begin{figure} % figuur 2
%\vspace{6pc}
\includegraphics[angle=270,width=0.9\textwidth]{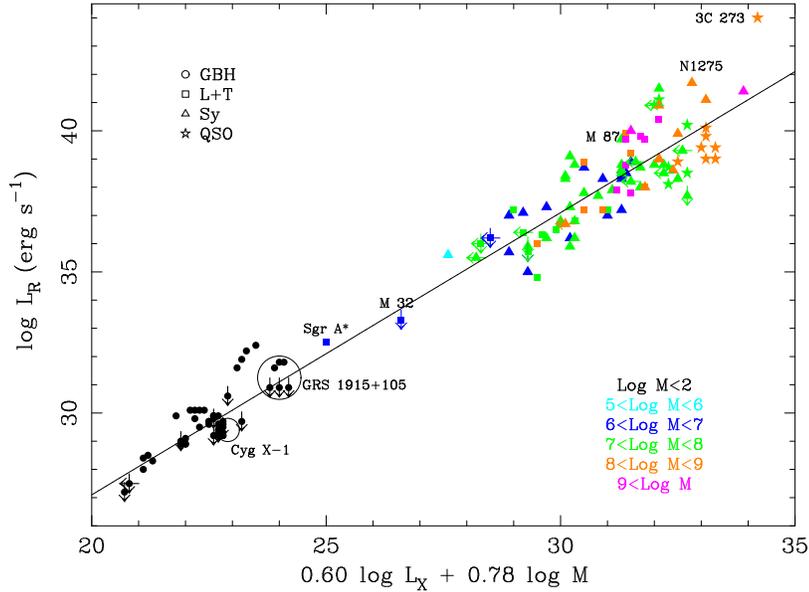}
\caption[]{The edge-on view of the ``fundamental plane of black hole
  activity''. The solid line shows the best fitting function (\ref{eq:fp}).}
\label{fig:fp}
\end{figure}
\end{center}

We can proceed to quantify the degree of correlation between our three
observables (radio luminosity at 5 GHz, $L_{\rm R}$, X-ray luminosity 
in the 2-10 keV band, $L_{\rm  X}$ and black hole
mass). We fit the data with the function $\log L_{\rm R}=\xi_{\rm RX}\log L_{\rm
  X} + \xi_{{\rm R}M}\log M + b_{\rm R}$, and obtain well constrained
values for the correlation coefficients, meaning that, 
if we define the instantaneous state of
activity of a black hole of mass $M$, by the
radio and hard X-ray luminosity of its
compact core, and represent such an object as a point in the
three-dimensional space ($\log L_{\rm R},\log L_{\rm X},\log M$), all 
black holes (either of stellar mass or supermassive)
will lie preferentially on a plane (the ``fundamental plane'' of
black hole activity, see Figure~\ref{fig:fp}), described by the following
equation:
\begin{equation}
\log L_{\rm R}=(0.60^{+0.11}_{-0.11}) \log L_{\rm X}
+(0.78^{+0.11}_{-0.09}) \log M + 7.33^{+4.05}_{-4.07}.
\label{eq:fp}
\end{equation}
with a dispersion $\sigma_{\rm R}=0.88$ (see Figure~\ref{fig:fp}). 

The value we obtain for the $\xi_{\rm RX}$ correlation coefficient is
consistent, within the errors, with that found in low/hard state GBH 
($\xi_{\rm RX} \approx 0.7$) by Gallo et al. (2003). 
This also means that individual GBH sources for which the
correlation between radio and X-ray luminosities is well established
(GX 339-4 and V404 Cyg) do indeed follow the same global trend defined
by black holes of all masses included in our sample.

\section{Scale invariance and jet dominance in black hole accretors}

In a recent paper, Heinz \& Sunyaev (2003) have
demonstrated that, under the general assumption that the jet formation
process is not qualitatively different among SMBH of different mass or
between SMBH and GBH, it is in fact possible to derive a universal
scaling between the jet (radio) luminosity at a given frequency
and both mass and accretion rate. The derived
relation is independent of the jet model and has scaling indices
that depend only on the (observable) spectral slope of the synchrotron
emission in the radio band ($\alpha_{\rm r}$), 
and on the accretion model. In particular,
it was shown that, if the magnetic field $B_0$ at the base of the jet is in
equipartition with the particles pressure and the total jet power
scales as $W_{\rm jet}\propto B_0^2 M^2 \propto \dot M$ (see also Falcke \&
Biermann 1995; Heinz et al., this proceedings), and
under the standard assumption that the electrons
responsible for the jet synchrotron emission follow a power-law
distribution in energy, with index $p=2$, then $L_{\rm R} \propto
M^{17/12-\alpha_{\rm r}/3}\dot m^{17/12+2\alpha_{\rm r}/3}$. For the
optically thick, flat spectrum cores we are mostly interested in 
($\alpha_{\rm r} \simeq 0$), then: 
\begin{equation}
L_{\rm R} \propto (M \dot m)^{1.42}.
\label{eq:hs03}
\end{equation}
To compare this prediction with the observed fundamental plane
correlation, one needs to explicate the dependence of the hard X-ray
luminosity on the accretion rate, $L_{\rm X}\propto \dot m^{q}$.
In MHD03 it was shown how indeed the prediction of standard synchrotron
theory for the radio-X-ray-mass correlation of Heinz \& Sunyaev (2003)
are verified if, and only if, 
$q \approx 2$, i.e. if the accretion process is radiatively 
inefficient. Detailed models for the X-ray emission from such flows do
indeed show that $q=2.3$ (MHD03), and in this case
$L_{{\rm R,}q=2.3} \propto \dot{m}^{1.38} M^{1.38} = \dot{M}^{1.38}$,
i.e., $L_{\rm R}$ scales with the physical accretion rate only.  This
is tantalizingly close to the predicted dependence of
eq.~(\ref{eq:hs03}), and such a result can be interpreted in the
following way: 1) the scale invariance hypothesis at the heart of the
Heinz and Sunyaev (2003) calculations is correct, and black  holes can
indeed be further unified by considering their masses and accretion
rates, and 2) the largest area of the fundamental plane is covered by
radiatively inefficient sources.

The total power released by the accretion/jet 
system may be written as
$W_{\rm tot} \simeq \dot M c^2 = L_{\rm bol}+W_{\rm jet}+W_{\rm
  adv,conv}$, where the first term on the right hand side is the total
radiated luminosity and the last one include contributions from
the energy advected and/or stored in the convective motions.
Our results suggest that the flow must be radiatively inefficient,
therefore, for small enough accretion rates we have
 $L_{\rm bol}\simeq \dot M \dot m c^2 \ll \dot M c^2 \sim W_{\rm jet}+W_{\rm
  adv,conv}$. On the other hand,  $ W_{\rm jet} \propto W_{\rm
  adv,conv} \propto  \dot M c^2$. Therefore, the issue of what
the relative fraction of the total accretion energy dissipated into
the jet is (or, alternatively, of when a source is ``jet dominated'';
Fender, Gallo \& Jonker  2003; Falcke, K\"ording \& Markoff 2004) 
reduces to the
determination of the value of the constant $W_{\rm jet}/W_{\rm adv,conv}$.
This requires the specification of a jet model or
the direct measure of the total kinetic power carried by the jet (see,
e.g. Heinz et al. 2004),
together with a dynamical model for the disc-jet coupling.

\section{Accretion mode changes}

It is well accepted, both from theory and observations, that accretion
can proceed in different modes (or states), with different radiative
efficiencies and spectral properties. By fitting the whole dataset at
our disposal with a single linear relationship (\ref{eq:fp}), we have
implicitly neglected the possibility of a global accretion mode
change. This is clearly implausible, as  the QSOs
and the bright Seyferts in our sample, which occupy the region of high
accretion rates, are independently known to have spectral
characteristics inconsistent with models of low radiative
efficiency. They should therefore depart from the observed correlations.
For GBHs, it has indeed been shown that the correlation between radio
and X-ray luminosity breaks down as the sources switch to their high
states (Maccarone 2003; Gallo, Fender and Pooley 2003). 
However, because both such modes of accretion are expected to occur
only above accretion rates about a few percent of Eddington, and
because another advective accretion mode is expected to ensue at around the
Eddington limit (due to efficient radiation trapping, see
e.g. Abramowicz et al. 1995), we would
expect the $\log M - \log L_{\rm R} - \log L_{\rm X}$ correlation to
break down only in a limited range of
$\dot m$. In other words, independently on the actual number of
radiatively efficient sources, the area of the fundamental plane covered
by them will always be limited, and the overall orientation of the
plane will always be dominated by the radiatively inefficient
ones. This also means that any (statistically significant) departure
from the fundamental plane relation could be used to identify
different modes of accretion. In fact, 
Maccarone, Gallo \& Fender (2003) have already shown how is
possible to use the fundamental plane relation as a baseline against
which identify AGN in a High/Soft State (HSS) analogous to that 
of X-ray binaries.

\begin{center}
\begin{figure} % figuur 1
%\vspace{6pc}
\includegraphics[angle=270,width=0.9\textwidth]{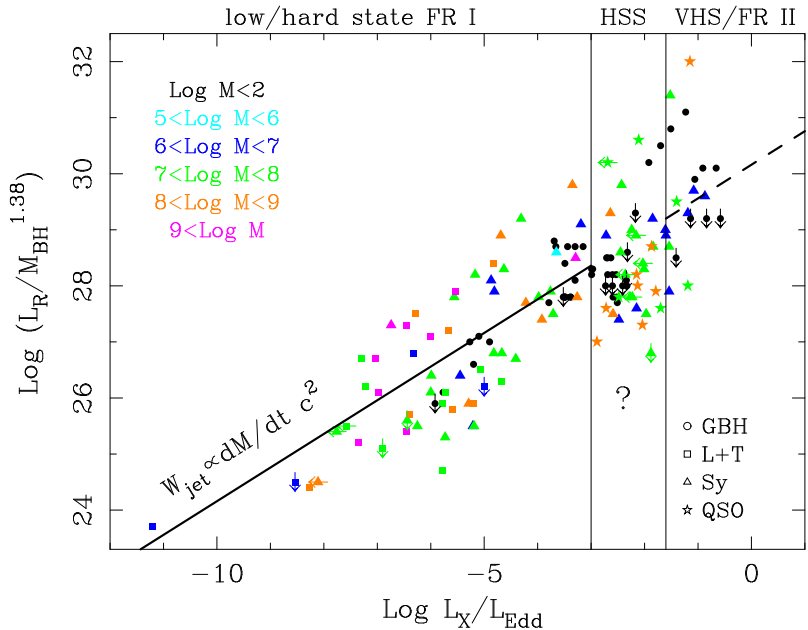}
\caption[]{The radio luminosity $\log L_{\rm R}$, divided by $M^{1.38}$ as a
function of the ratio $L_{\rm X}/L_{\rm Edd}$.   Upper
  limits are marked with arrows, different symbols indicate objects
  belonging to different spectral classes and different colors objects
  in different mass bins. Two vertical lines mark
the boundary of the region where we expect the critical luminosity for the
mode change between radiatively inefficient and efficient
  accretion. To the left of these lines, GBH are in the low/hard state
  and SMBH are mostly FR I. To the right, GBH are in the
  Very-High state (VHS) and SMBH are mostly FR II. In between,
  possibly the restricted region of the parameter space where pure
  discs accretion is allowed (High/Soft State, HSS).}
\label{fig:lr_mdotx_fit}
\end{figure}
\end{center}

The analogy with GBH can indeed be very useful to classify and understand the
properties of bright AGN. Let us consider, for
example, a recent paper by Fender,
Belloni and Gallo (2004), where it was shown that the intermittent, powerful
radio flares from the so-called microquasars, associated with the
rapid variability in the Very High State (VHS), seem to follow a
similar radio-X-ray correlation as the low/hard state sources, albeit
with a much larger scatter. Can we extend this result to the larger
family of radio loud AGN? 

In Figure~\ref{fig:lr_mdotx_fit} we plot, as a function of
the ratio $L_{\rm X}/L_{\rm Edd}$, the radio luminosity divided by
$M^{1.38}$ of all the sources in our sample.  
As expected, by rescaling the radio luminosity in such
a way all the different tracks corresponding to different mass bins in
Fig.~\ref{fig:lr_mdot2} collapse into a single one, with some
residual scatter.  The region between the two vertical lines
corresponds to the theoretically 
expected values of $L_{\rm X}/L_{\rm Edd}$ above
which a change of accretion mode, from radiatively inefficient to
standard radiatively efficient is expected to occur. 
To the left of these lines, GBH are in the low/hard state
and SMBH are mostly FR I, or low-luminosity, radio loud AGN. 
To the right, GBH are in the Very-High state (VHS) 
and SMBH are mostly FR II. In between,
possibly lies the restricted region of the parameter space where pure
discs accretion is allowed (HSS). Interestingly, it appears as if both
low and high luminosity sources at the two sides of the HSS region
obey a similar scaling $W_{\rm jet} \propto \dot M c^2$.

We may thus speculate that the famous (and still much debated) 
radio loud/radio quiet dichotomy of
quasars will appear only at the highest values of $\dot m$, and be
caused mainly by a switch of accretion mode analogous to the high/very
high transition in GBH.  At low accretion rates, black holes seem to
follow the more regular behavior circumscribed by the fundamental
plane of eq.~(\ref{eq:fp}). Such sources not only tend to be radio
loud (Ho \& Peng 2001; Ho 2002), but also their radio loudness parameter,
$R_{\rm X}$ (here defined as the ratio of radio to X-ray luminosity),
obeys the following scaling: $R_{\rm X}\equiv L_{\rm R}/L_{\rm X}
\propto L_{\rm X}^{-2/5} M^{4/5}$. Therefore, the smaller the 
X-ray luminosity, the more radio loud these sources are. 
In this regime, no dichotomy need be expected, as
already suggested by Nagar et al. (2002). 

\section{Conlcusions}
We have argued that the fundamental plane analysis presented here is a
powerful tool to extend the unified scheme of accreting black
holes. Such a relation between mass, radio and hard X-ray luminosity
is affected very little by obscuration and beaming, provided that
sources whose relativistic jets are in our line of sight can be
effectively identified and excluded.  Thus, the fundamental plane
relation does not depend on orientation, and as such is complementary
to the standard unification scheme. Moreover, the relation itself is
perfectly consistent with the scaling relations predicted by standard
synchrotron theory under a scale invariance assumption. The main
scaling parameters are the mass of the black hole and its accretion
rate. Finally, we have shown how the observed correlation can be
effectively used to classify objects on the basis of their mode of
accretion (and/or accretion/ejection coupling) rather than just on
specific observational characteristics, as in the true spirit of
unification models.

\end{article}

\begin{thebibliography}{}

\bibitem[\protect\citeauthoryear{Abramowicz et al.}{1995}]{abr95}
Abramowicz, M.~A., X. Chen, S. Kato, J.-P. Lasota, O. Regev, 1995,
ApJ, 438, L37 

\bibitem[\protect\citeauthoryear{Begelman et al.}{1984}]{bbr84}
Begelman, M.~C., R.~D. Blandford and M.~J. Rees, 1984,
Rev. Mod. Phys., 56, 255

\bibitem[\protect\citeauthoryear{Falcke and Biermann}{1995}]{fb95}
Falcke, H. and P.~L. Biermann, 1995, A\&A, 293, 665

\bibitem[\protect\citeauthoryear{Falcke et al.}{2004}]{fkm04}
Falcke, H., E. K\"ording and S. Markoff, 2004, A\&A, 414, 895

\bibitem[\protect\citeauthoryear{Fender et al.}(2003)]{fgj03}
Fender, R.~P., E. Gallo and P. Jonker, 2003, MNRAS, 343, L99 

\bibitem[\protect\citeauthoryear{Fender et al.}(2004)]{fbg04}
Fender, R.~P., T. Belloni and E. Gallo, 2004, MNRAS in press. astro-ph/0409360

\bibitem[\protect\citeauthoryear{Franceschini et al.}{1998}]{fvf98}
Franceschini, A., S. Vercellone, A.~C. Fabian, 1998, MNRAS, 297, 817

\bibitem[\protect\citeauthoryear{Gallo et al.}{2003}]{gfp03}
Gallo, E., R.~P. Fender and G.~G. Pooley, 2003, MNRAS, 344, 60

\bibitem[\protect\citeauthoryear{Heinz and Sunyaev}{2003}]{hs03}
Heinz, S. and R.~A. Sunyaev, 2003, MNRAS, 343, L59

\bibitem[\protect\citeauthoryear{Heinz et al.}{2004}]{h04}
Heinz, S., R.~A. Sunyaev, A. Merloni, T. Di Matteo, 2004, to appear in
the Proceedings of "Growing Black Holes", Garching, Germany, June 21-25, 2004. Eds. A. Merloni, S. Nayakshin and R. Sunyaev, Springer-Verlag series of "ESO Astrophysics Symposia".

\bibitem[\protect\citeauthoryear{Ho}{2002}]{ho02}
Ho, L.~C., 2002, ApJ, 564, 120

\bibitem[\protect\citeauthoryear{Ho and Peng}{2001}]{hp01}
Ho, L.~C. and C.~Y. Peng, 2001, ApJ, 555, 650

\bibitem[\protect\citeauthoryear{Kormendy and Richstone}{1995}]{kr95}
Kormendy, J. and D. Richstone, 1995, ARA\&A, 33, 581


\bibitem[\protect\citeauthoryear{Maccarone}{2003}]{mac03}
Maccarone, T., 2003, A\&A, 409, 697

\bibitem[\protect\citeauthoryear{Maccarone et al.}{2003}]{mgf03}
Maccarone, T., E. Gallo and R.~P. Fender, 2003, MNRAS, 345, L19

\bibitem[\protect\citeauthoryear{Magorrian et al.}{1998}]{mag98}
Magorrian, J. et al., 1998, AJ, 115, 2285 

\bibitem[\protect\citeauthoryear{Merloni et al.}{2003}]{mhd03}
Merloni, A., S. Heinz and T. Di Matteo, 2003, MNRAS, 345, 1057 (MHD03)

\bibitem[\protect\citeauthoryear{Nagar et al.}{2002}]{nag02}
Nagar N.~M., A.~S. Wilson, H. Falcke, J.~S. Ulvestad, 2002, A\&A, 392, 53


\bibitem[\protect\citeauthoryear{Rawlings and Sanders}{1991}]{rs91}
Rawlings, S. and R. Saunders, 1991, Nature, 349, 138

\bibitem[\protect\citeauthoryear{Urry and Padovani}{1995}]{up95}
Urry, C.~M. and P. Padovani, 1995, PASP, 107, 803


\end{thebibliography}
\end{document}